\documentclass[final,5p,times,twocolumn]{elsarticle}
\usepackage{graphicx}
\usepackage{amssymb}
\usepackage{amsmath}
\usepackage[margin=0pt,font=small,labelfont=bf,labelsep=colon]{caption}
\usepackage{algorithm}
\usepackage{algpseudocode}
\usepackage{listings}
\usepackage{color}
\usepackage{textcomp}
\usepackage{url}
\usepackage{lineno}

\biboptions{sort&compress}
\journal{Carbon}
\begin{document}
\begin{frontmatter}
\title{First principles investigation of nitrogenated holey graphene}
\author{Cui-Yan Xu $^{1}$}
\author{Hai-Kuan Dong $^{2}$ \corref{cor2}}
\ead{dhk@bhu.edu.cn}
\author{Li-Bin Shi $^{2}$ \corref{cor2}}
\ead{slb0813@126.com}
\cortext[cor1]{Corresponding authors}
\address{$^{1}$College of Science, Liaoning University of Technology,  Liaoning Jinzhou 121000, China}
\address{$^{2}$School of Mathematics and Physics, Bohai University, Liaoning Jinzhou 121013, China}
\begin{abstract}
Nitrogenated holey graphene (NHG) has attracted much attention because of its semiconducting properties. However, the stacking orders and defect properties have not been investigated. In this letter, the structural and stacking properties of NHG are first investigated. We obtain the most stable stacking structure. Then, the band structures for bulk and multilayer NHG are studied. Impact of the strain on the band gaps and bond characteristics is discuss. In addition, we investigate formation mechanism of native defects of carbon vacancy ($\rm V_C$), carbon interstitial ($\rm C_i$), nitrogen vacancy ($\rm V_N$), and nitrogen interstitial ($\rm N_i$) in bulk NHG. Formation energies and transition levels of these native defects are assessed.
\end{abstract}
\end{frontmatter}
\section{Introduction}
\label{Introduction}
With the development of science and technology, the search for new functional materials has been an important goal of scientific research \cite{he2011integrations}. Because the excellent properties of two-dimensional (2D) materials meet the requirements of future nanoelectronics, they have attracted much attention \cite{Li2015First,Shi2017First}. Recently, a unique two-dimensional nanosheet of nitrogenated holey graphene (NHG) with a stoichiometry of $\rm C_2N$ in the plane was synthesized by a simple wet-chemical reaction in 2015 \cite{Mahmood2015Nitrogenated}.
\par
Graphene has a zero band gap, which limits its application in the field of electronic devices \cite{Zhou2009Ferromagnetism}. NHG is superior to graphene in many properties. First of all, it is a semiconductor with direct band gap \cite{Tromer2017Atomic}. Another advantage of NHG over graphene is the low cost synthesis \cite{Mahmood2015Nitrogenated}. In addition, the electronic and optical properties of NHG can be tuned by a large range strain \cite{Guan2015Effects}. As a result, it becomes a potential candidate for nano-electronics, optoelectronics, and solar driven photocatalysis. Previously, some studies on NHG have been carried out  \cite{Sahin2015Structural,Zhang2016Giant,Mortazavi2016Thermal,Rahaman2017A,Yagmurcukardes2016Nitrogenated,Kang2015Heterostructures,Yang2017Tuning}, which includes atomic adsorption, thermal conductivity, mechanical characteristics, and magnetic properties. However, the stacking orders and native defects are still paid little attention.
\par
Here we try to shed light on structural characteristics of NHG. First of all, in order to find the most stable structure we study the stacking characteristics of NHG. Then, we investigate the band structures for bulk and multilayer NHG. Impact of the strain on the band gaps and bond characteristics is investigated. At last, we investigate the formation energies and ionization characteristics of native defects in bulk NHG structure. We discuss which kinds of defects are most likely to form in NHG.
\section{Models and Methods}
The calculations on structural and electronic properties of NHG are performed using the CASTEP code, which is based on density functional theory and plane-wave pseudopotential method. Generalized gradient approximation (GGA) of Perdew, Burke, and Ernzerhof (PBE) is chosen as the exchange-correlation functional. We also attempt to improve the band gap using hybrid Heyd-Scuseria-Ernzerh functional(HSE) \cite{heyd2003hybrid}. The monolayer NHG is a single 2D sheet of atoms with uniform periodic holes in a fused aromatic network structure \cite{Tromer2017Atomic}. The orthorhombic unit cell consists of 36 atoms, including of 12 nitrogen atoms and 24 carbon atoms. In order to correctly describe the interlayer interaction, Van der Waals (VdW) interaction between the NHG layers is calculated by Grimme¡¯s DFT-D correction \cite{Grimme2006}. A vacuum region of 15 \r{A} is used to avoid the interaction between the top and bottom layers. The ion core electrons are replaced by norm-conserving pseudopotentials. The valence electron configurations for C and N are considered as $2s^22p^2$ and $2s^22p^3$, respectively. The convergence tests of the total energy with respect to the plane wave energy cutoff and k-points sampling have been carefully examined. The final set of the energy cutoff is 700 eV, and the integration is performed by $3\times2\times1$ k point sampling, which is generated by the Monkhorst-Pack scheme. All atoms in unit cell are fully relaxed. The optimization will finish if the force convergence criterion of 0.01 eV/\r{A} is satisfied.
\section{Results and Discussion}
Figure 1 presents crystal structure for NHG. The crystal is composed of evenly distributed holes. The nitrogen and carbon atoms are located in a flat structure. The lattice constants of NHG are a= 8.37 \r{A} and 14.44 \r{A}, which is in agreement with previous investigation of a= 8.33 \r{A} and 14.33 \r{A} \cite{Tromer2017Atomic}. The bond lengths in NHG shown in Fig. 1 are 1.43 \r{A}, 1.48 \r{A} for the C-C(1), C-C(2), and  1.35 \r{A}, 1.34 \r{A} for the C-N(1), C-N(2), which is in accordance with the previous result \cite{Zhu2017Half}. The bond angle shown in Fig. 1 is $\alpha$=116.54$^\circ$. The calculations also show the graphene is a zero band gap semiconductor with C-C bond length of 1.42 \r{A} and bond angle of $\alpha$=120$^\circ$. Figure 2 shows the partial density of states for NHG structure. It is found from Fig. 2 that the valence band is mainly composed of C 2s, 2p and N 2s, 2p, while the conduction band is made up of C 2p and N 2p. The upper valence band is mainly determined by C 2p and N 2p, while the lower valence band is mainly determined by C 2s and N 2s. It is observed that the C 2s, 2p hybridizes with N 2s, 2p below Fermi level. Figure 3 shows the charge density difference of NHG crystal structure. The charge density is mainly distributed between C and N, which strongly suggests that the covalent bond bond is formed between C and N. It is observed that the holey site is surrounded by negatively charged N atoms, which is in agreement with previous investigation \cite{Sahin2015Structural}. Mulliken atomic populations show that each C atom donates 0.17 e, and each N atom obtain 0.34 e.
\begin{figure}
\centering
\includegraphics[width=8.5 cm]{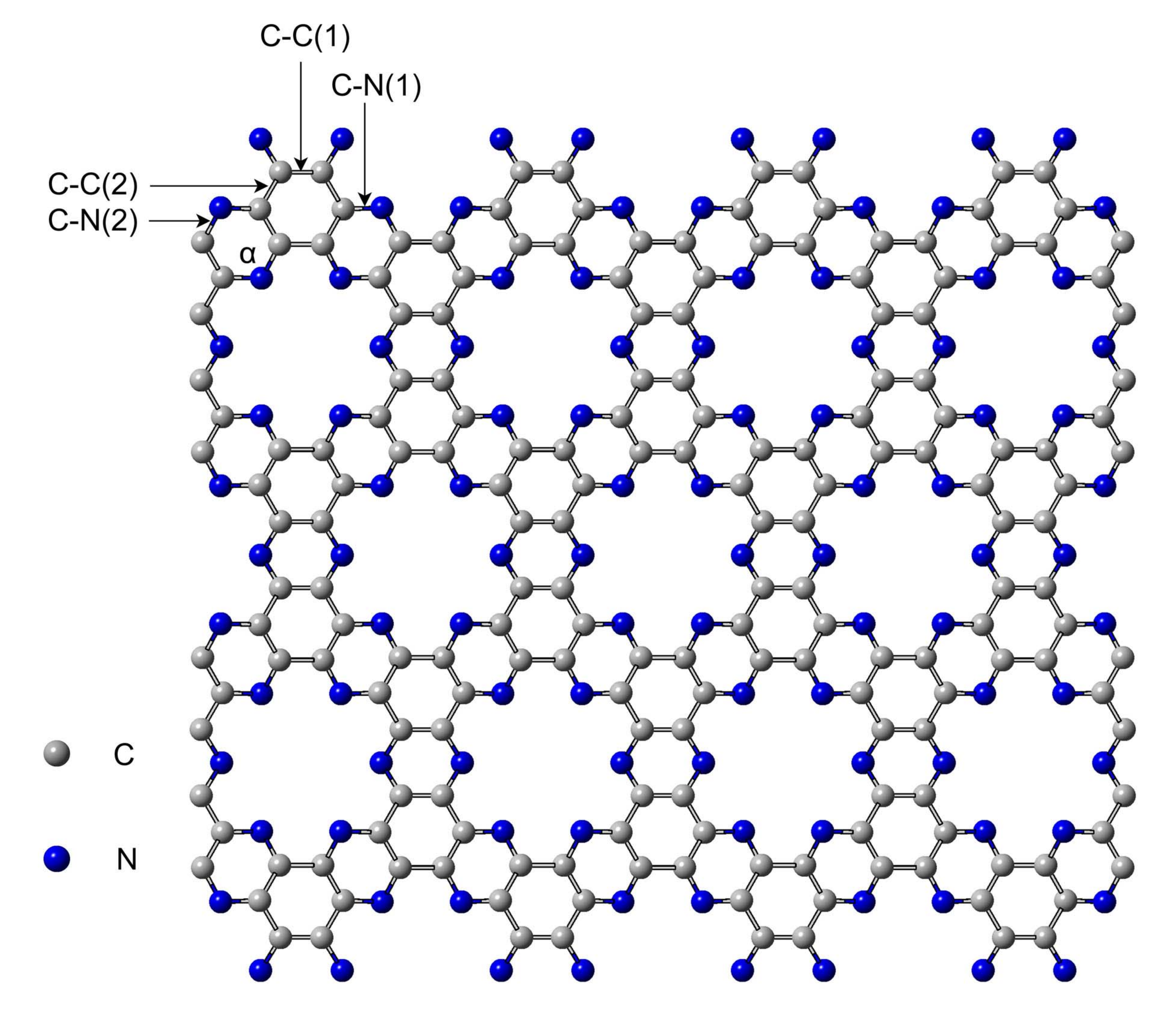}
\caption{NHG crystal structure composed of nitrogen (blue) and carbon (gray) atoms. Some specific bond lengths (C-C(1),C-C(2), C-N(1) and C-N(2)) and bond angles ($\alpha$) are marked in the graph}
\label{figure1}
\end{figure}

\begin{figure}
\centering
\includegraphics[width=8.5 cm]{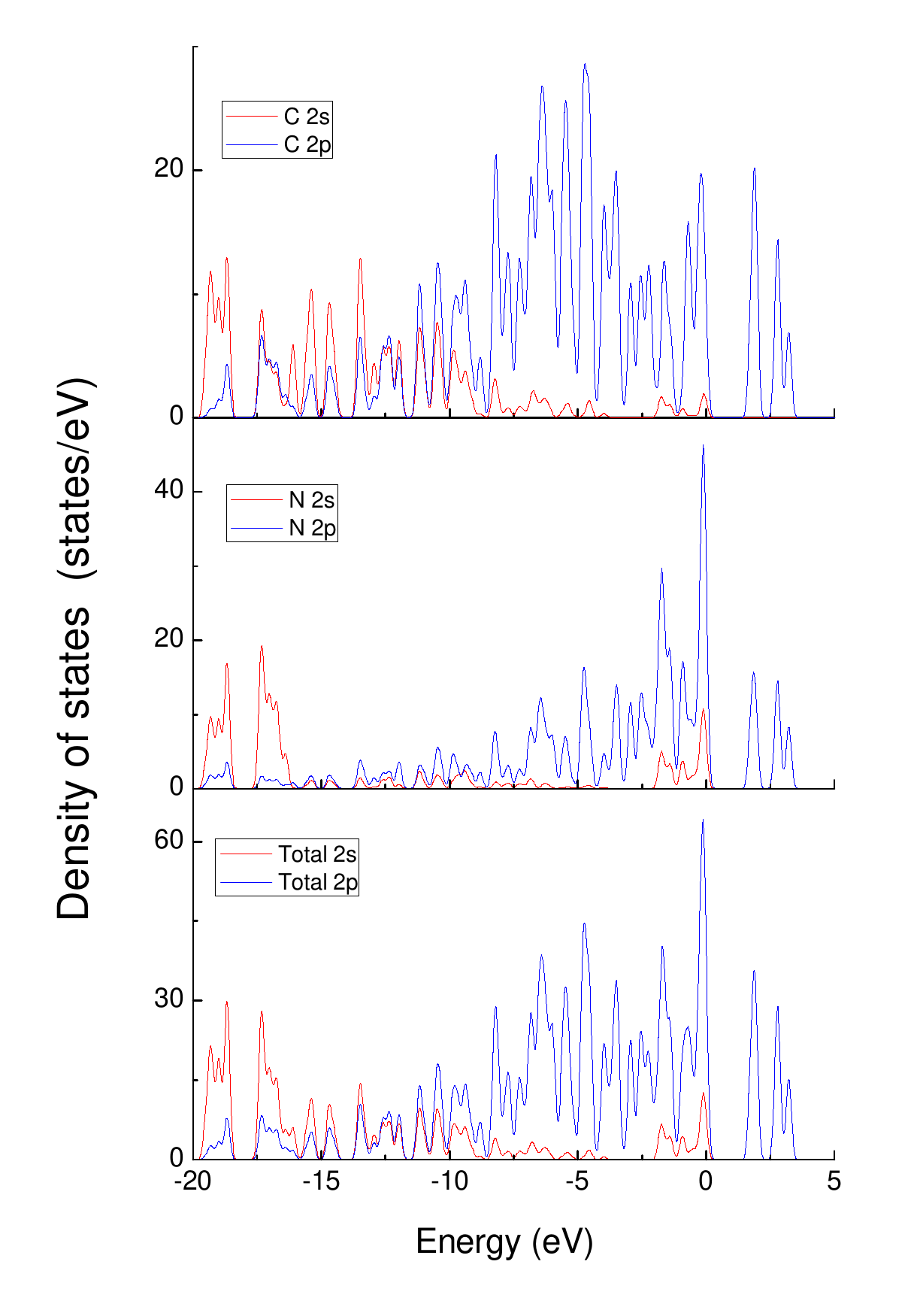}
\caption{The partial density of states for NHG structure. The Fermi level is set to zero. The calculation is carried out in $2\times1\times1$ supercell. }
\label{figure2}
\end{figure}

\begin{figure}
\centering
\includegraphics[width=7 cm]{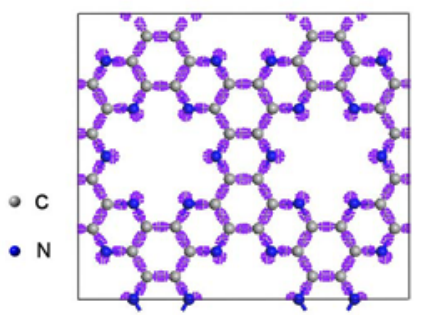}
\caption{The electron density difference distributions for NHG structure. The calculation is carried out in $2\times1\times1$ supercell.}
\label{figure3}
\end{figure}

\par
In this study, seven typical orthorhombic unit cell structures of A, B, C, D, E, F, and G are chosen for NHG, which is shown in Fig. 4. Stacking orders of A-A, A-B, A-C, A-D, A-E, A-F, and A-G can almost exhaust all conceivable stacking configurations of NHG. A-A has the same atomic configurations on both top and bottom layers, while A-B, A-C, A-D, A-E, A-F, and A-G consider the possibility of various atomic configurations at the center of the hole. To explore the stacking stability of bilayer NHG, we calculate the binding energies as a function of interlayer spacing (see Fig. 4). The binding energy is defined as the total energy difference per atom between bilayer NHG and the two single monolayers. The interlayer spacing and binding energy are 3.6 \r{A} and -9.69 meV/atom for A-A stacking, 3.3 \r{A} and -20.20 meV/atom for A-B stacking, 3.2 \r{A} and -23.99 meV/atom for A-C stacking, 3.1 \r{A} and -18.34 meV/atom for A-D stacking, 3.1 \r{A} and -24.81 meV/atom for A-E stacking, 3.1 \r{A} and -26.38 meV/atom for A-F stacking, and 3.3 \r{A} and -18.33 meV/atom for A-G stacking. The calculation also shows that the graphene with the interlayer spacing of 3.2 \r{A} and interfacial binding energy of -57.53 meV/atom. The binding energies are similar to previous investigations of 2D materials of $\rm MoS_2$ \cite{2014Stacking}, and less than those of $\rm CoO/MnO$ \cite{Yao2008first} and $\rm Mg/Al_4C_3$ \cite{li2013first} interfaces. This result indicates that the interlayer interaction is weak for NHG. Among these relatively stable structures, A-F stacking is more stable than other stacking orders. Coulomb interaction between monolayers may play an important role in distinguishing the stability between the stacking orders.
\begin{figure}
\centering
\includegraphics[width=8.5 cm]{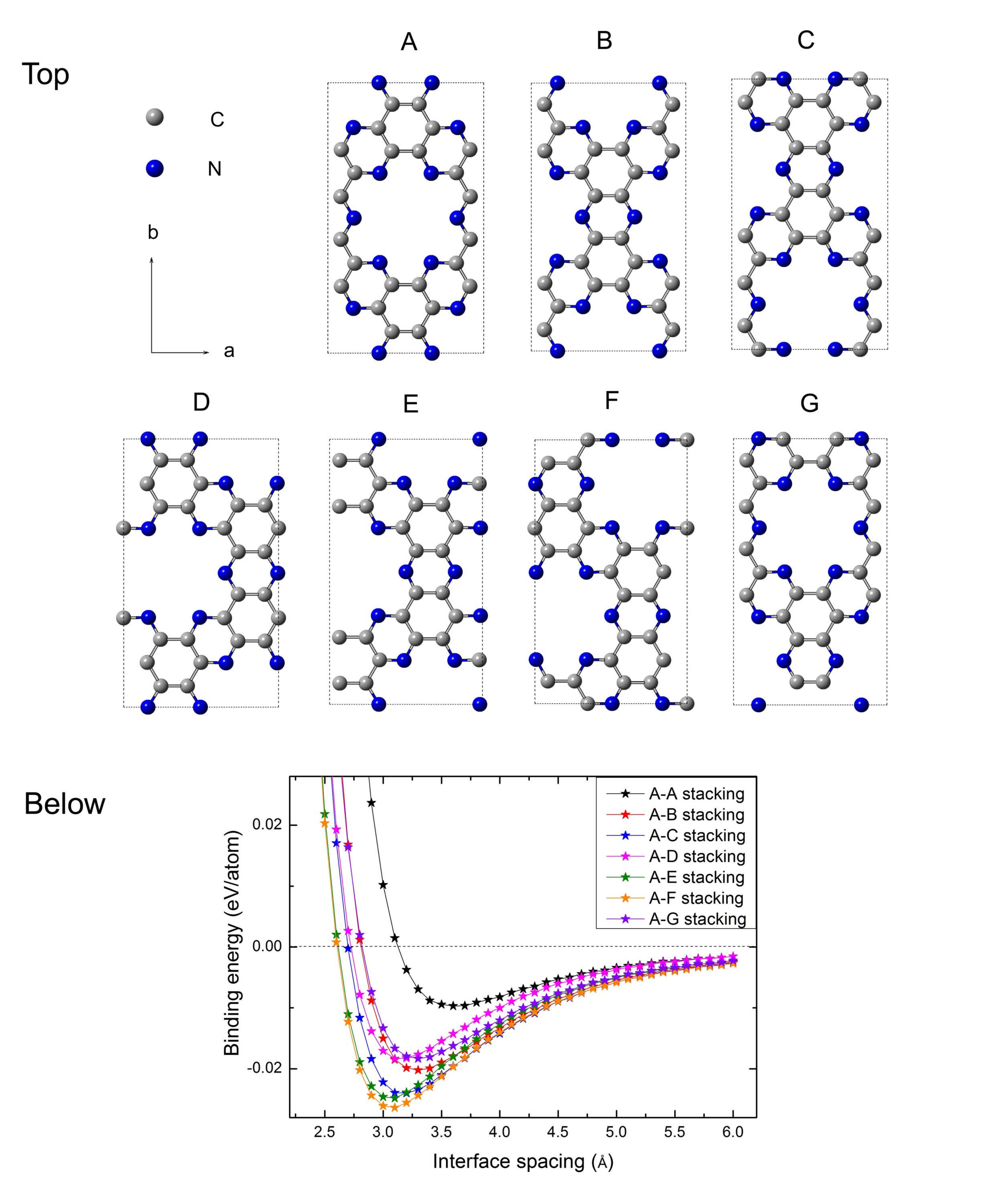}
\caption{Top: Seven typical orthorhombic unit cell structures of A, B, C, D, E, F, and G for NHG crystal; Below: Interfacial binding energies as a function of interlayer spacing.}
\label{figure4}
\end{figure}

\par
Figure 5 exhibits the band structures for bulk and different thickness layers of $\rm NHG$ by GGA-PBE and HSE functionals. The band structures suggest that $\rm NHG$ is a semiconductor with direct band gap. It is noted that the band gap decreases with the increase of the number of $\rm NHG$ layers. The GGA-PBE and HSE band gaps are 1.66 eV and 2.34 eV for monolayer $\rm NHG$, 1.52 eV and 2.12 eV for 2-layer $\rm NHG$, 1.44 eV and 2.02 eV for 3-layer $\rm NHG$, 1.36 eV and 1.94 eV for bulk $\rm NHG$.  The GGA-PBE functional usually underestimates the interaction between electrons in semiconductors or oxides, leading to underestimation of band gap \cite{2017First}. In order to improve the band gap, the interaction between electrons must be enhanced. The hybrid Heyd-Scuseria-Ernzerhof functional (HSE) combines screened Hartree Fock exchange with the GGA of Perdew, Burke, and Ernzerhof (PBE) \cite{heyd2003hybrid,heyd2004efficient,paier2006erratum,heyd2005energy}, which can enhance interaction between electrons, and improve the band gaps of semiconductors or oxides \cite{lyons2011the,dewalle2013defects,choi2013native}. Comparing with experimental value of 1.96 eV for the monolayer $\rm NHG$ \cite{Mahmood2015Nitrogenated}, GGA-PBE functional underestimates the band gap, while HSE functional overestimates the band gap.

\begin{figure}
\centering
\includegraphics[width=8.5 cm]{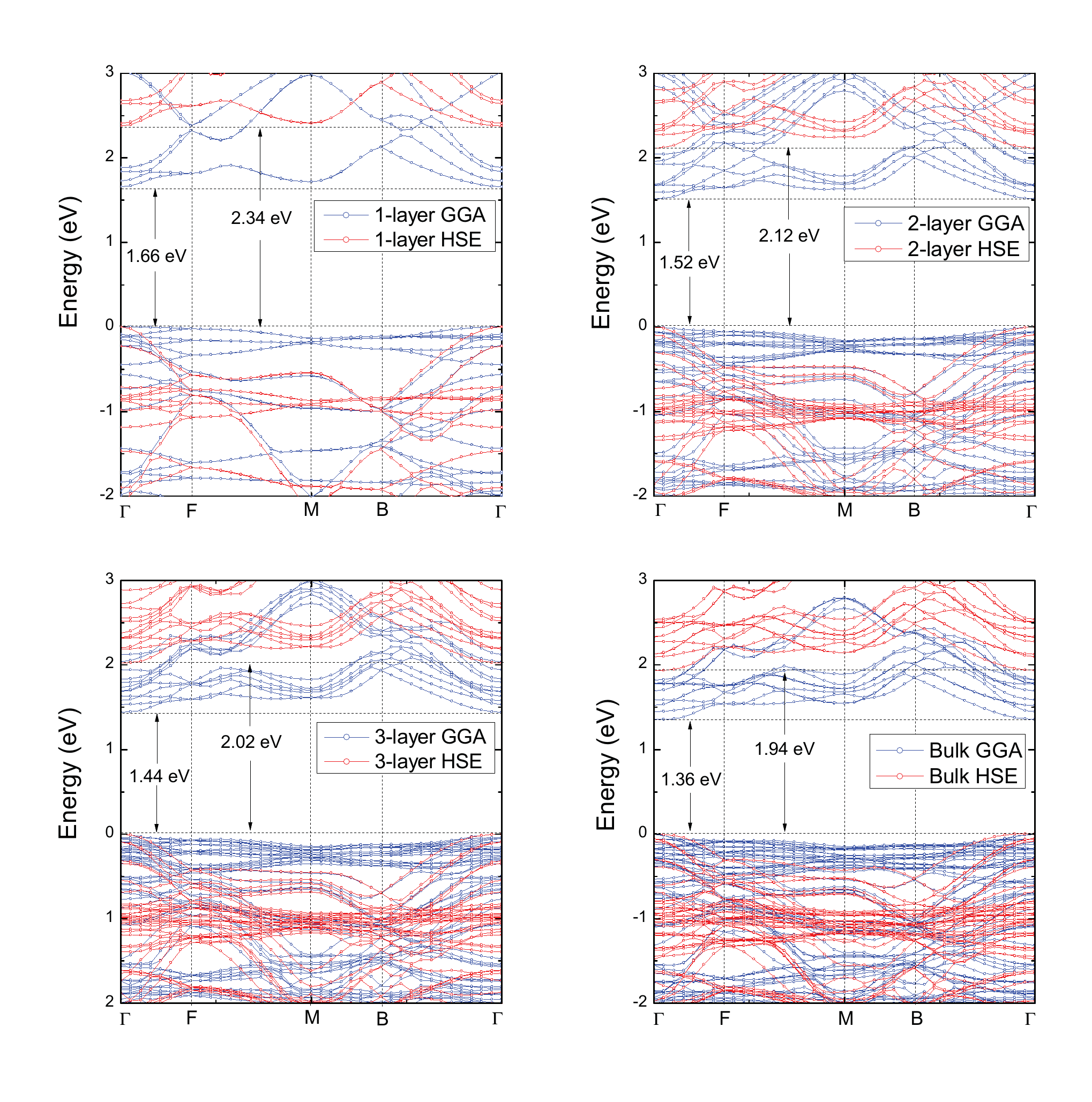}
\caption{Band structures for bulk and different thickness layers of $\rm NHG$. The blue and red point connections represent GGA-PBE and HSE results. The arrows indicate the band gap values calculated by GGA-PBE and HSE functionals.}
\label{figure5}
\end{figure}

\par
Strain engineering can improve physical performances of semiconductors, in particular, their transport properties \cite{liu2015strain,yu2015phase,chang2015hole,zhang2009physical}. For monolayer NHG, we investigate the strain behavior by GGA-PBE functional. The lattice strain is defined as $\varepsilon= \frac{c-c_0}{c_0}\times 100\%$ \cite{kaloni2013hole}, where $c$ and $c_0$ are lattice constants of strained and unstrained $\rm NHG$. The lattice strain is set in the range from -6\% to 6\% with an interval of 2\%.  Table 1 represents band gap, valence band maximum, and bond lengths under a, b uniaxial and ab biaxial strain. Figure 6 presents the band gaps under different lattice strain. The direct band gap varies in the range from 1.297 eV - 1.708 eV as the strain is applied to the monolayer $\rm NHG$. It is observed from Fig. 06 that the band gap reaches its maximum as strain $\varepsilon= 2\%$. The results show that the carrier transport properties can be modulated by the lattice strain. Previously, Guan et al. investigated effects of strain on electronic and optic properties of NHG \cite{Guan2015Effects}. It is observed that the NHG remains a direct-gap semiconductor under biaxial strain of $\varepsilon$=4\%, 8\% and 12\%. They found that the band gap reaches its maximum as biaxial strain is about $\varepsilon$=8\%. By comparing with the previous investigation \cite{Guan2015Effects}, it is found that the band gap reaches the maximum under a certain biaxial tensile strain. The difference may be related to the exchange-correlation functional chosen by the calculation. The valence band maximum decreases under the compressive strain. The calculated results also show that the bond lengths of N-C and C-C will increase with the strain changing from compressive strain to tensile strain. This change in the bond length indicates that NHG exhibits very good elastic characteristics, which is in agreement with previous investigation \cite{Mortazavi2016Thermal,Rahaman2017A}.

\par
\begin{figure}
\centering
\includegraphics[width=8.5 cm]{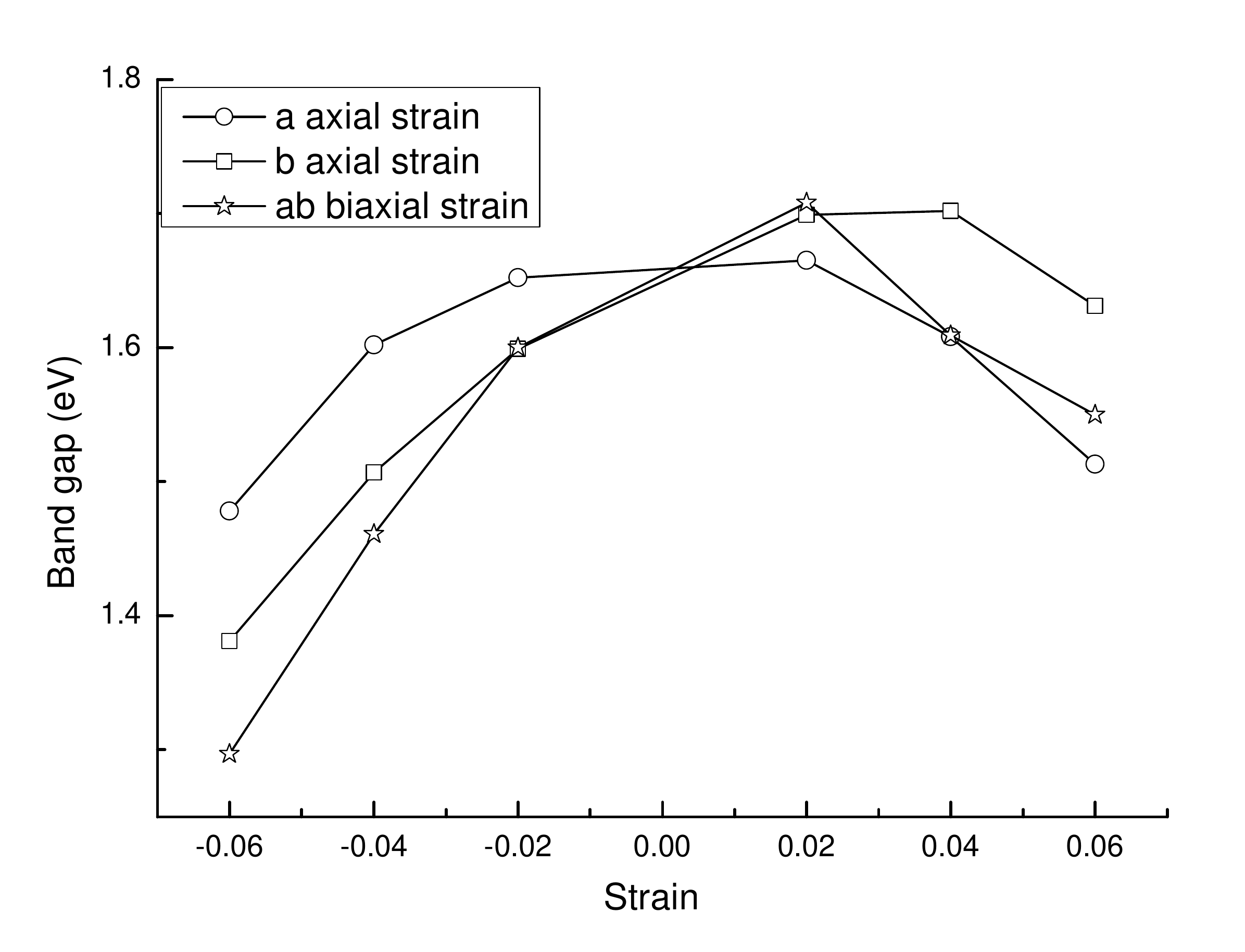}
\caption{Band gaps under different strain for monolayer NHG.}
\label{figure6}
\end{figure}

\begin{table*}
\linespread{1.3}
\centering
\caption{Band gap, valence band maximum, and bond lengths of $\rm C-C(1)$, $\rm C-C(2)$, $\rm C-N(1)$ and $\rm C-N(2)$.}
  \label{tb2}
  \scriptsize
  \begin{tabular}{c|ccccccc}
    \hline
    \hline
    Strain              &Type    &$\rm Band gap(eV)$    &$\rm VBM (eV)$   &$\rm C-N(1)(\r{A})$  &$\rm C-N(2)(\r{A})$ &$\rm C-C(1)(\r{A})$  &$\rm C-C(2)(\r{A})$ \\
    \hline
    $\varepsilon$=-6\%  &ab axis     &1.297         &5.634       &1.321      &1.305     &1.376   &1.388 \\
                        &a axis      &1.478         &5.858       &1.324      &1.324     &1.406   &1.445 \\
                        &b axis      &1.381         &5.734       &1.349      &1.320     &1.395   &1.413 \\
    $\varepsilon$=-4\%  &ab axis     &1.461         &5.819       &1.332      &1.317     &1.393   &1.415 \\
                        &a axis      &1.602         &5.998       &1.335      &1.328     &1.413   &1.455 \\
                        &b axis      &1.507         &5.881       &1.351      &1.327     &1.405   &1.432 \\
    $\varepsilon$=-2\%  &ab axis     &1.600         &5.986       &1.343      &1.328     &1.411   &1.444 \\
                        &a axis      &1.652         &6.059       &1.345      &1.333     &1.420   &1.465 \\
                        &b axis      &1.599         &5.993       &1.352      &1.333     &1.416   &1.453 \\
    $\varepsilon$=2\%   &ab axis     &1.708         &6.140       &1.364      &1.348     &1.439   &1.509 \\
                        &a axis      &1.665         &6.076       &1.362      &1.344     &1.430   &1.485 \\
                        &b axis      &1.699         &6.130       &1.355      &1.343     &1.433   &1.498 \\
   $\varepsilon$=4\%    &ab axis     &1.609         &6.108       &1.373      &1.358     &1.452   &1.548 \\
                        &a axis      &1.608         &6.100       &1.370      &1.349     &1.437   &1.495 \\
                        &b axis      &1.702         &6.191       &1.357      &1.349     &1.441   &1.525 \\
   $\varepsilon$=6\%    &ab axis     &1.550         &6.076       &1.382      &1.367     &1.464   &1.593 \\
                        &a axis      &1.513         &6.057       &1.379      &1.354     &1.444   &1.506 \\
                        &b axis      &1.631         &6.207       &1.359      &1.354     &1.447   &1.554 \\
    \hline
    \hline
  \end{tabular}
\end{table*}

\par
The defect concentration is inversely proportional to exponential functional of its formation energy \cite{janotti2007native}. The defect formation energy can be obtained from the total energy calculation \cite{li2006design,kohan2000first,limpijumnong2004doping,huang2008magnetism}.
 \begin{equation}
 \Delta{E_{\rm f}(\alpha,q)}=\Delta{E(\alpha,q)}+\sum_{\alpha}n_{\alpha}{\Delta\mu_{\alpha}}+qE_{\rm f}
 \end{equation}
 \begin{equation}
 \Delta{E(\alpha,q)}=E(\alpha,q)-E(\rm C_2N)+\sum_{\alpha}n_{\alpha}\mu_{\alpha}+\textit{q}\textit{E}_{\rm v}
 \end{equation}
where $E(\alpha, q)$ is total energy for bulk $\rm C_2N$ containing the relaxed defect $\alpha$ in the charge state $q$. $E(\rm C_2N)$ is the total energy for the same bulk $\rm C_2N$ in the absence of defect. $n_{\rm \alpha}$ is the number of the defects. $E_{\rm v}$ represents the energy of the valence band maximum (VBM). $\mu_{\rm \alpha}$ is the chemical potential of the element $\alpha$. $\Delta \mu_{\rm \alpha}$ is the relative chemical potential. $\Delta \mu_{\rm \alpha}$+$\mu_{\rm \alpha}$ is the absolute value of the chemical potential of atom $\alpha$. We investigate formation mechanism of carbon interstitial ($\rm C_i$), carbon vacancy ($\rm V_C$), nitrogen interstitial ($\rm N_i$), and nitrogen vacancy ($\rm V_N$) in bulk NHG. It is found that $\rm C_i$ and $\rm N_i$ is easy to form in hole for nitrogenated holey graphene. For the neutral charge state, $\rm C_i$ is easy to combine with two N atoms with bond length of 1.46 \r{A}, while $\rm N_i$ only combines with one N atom with bond length of 1.30 \r{A}. $\rm C_i$ and $\rm V_N$ in formation energy is lower under carbon rich limit, while $\rm N_i$ and $\rm V_C$ in formation energy is lower under nitrogen rich limit. The formation energy of native defects under carbon rich limit is presented in Fig. 7. The slope corresponds to the charge state $q$ used in Eq. (1). Change in the slope indicates transition in the charge states. It is found that the neutral states of these defects are unstable. Similar results are also found on $\rm TO_2$ \cite{peng2008First} and  $\rm ZnO$\cite{janotti2007native}. The charge states of these defects are mainly determined by +1, -1, and -4 for main Fermi level range. For $\rm C_i$, the major transition levels are $\varepsilon$(+4/+3)=$E_v$+0.15 eV, $\varepsilon$(+1/-1)=$E_v$+0.78 eV, and $\varepsilon$(-3/-4)=$E_c$-0.19 eV. For  $\rm N_i$, there are $\varepsilon$(+3/+1)=$E_v$+0.23 eV, $\varepsilon$(+1/-1)=$E_v$+1.06 eV and $\varepsilon$(-3/-4)=$E_c$-0.18 eV. For $\rm V_C$, the major transition levels are $\varepsilon$(+3/+1)=$E_v$+0.20 eV, $\varepsilon$(+1/-1)=$E_v$+0.62 eV and $\varepsilon$(-3/-4)=$E_c$-0.30 eV. For $\rm V_N$, the major transition levels are $\varepsilon$(+3/+1)=$E_v$+0.13 eV, $\varepsilon$(+1/-1)=$E_v$+0.36 eV, $\varepsilon$(-1/-2)=$E_c$-0.59 eV, and $\varepsilon$(-3/-4)=$E_c$-0.32 eV. The result shows that the defects act as carrier traps. In addition, charge defects also lead to region of fixed charge that can scatter carrier and shift the threshold voltage of the device \cite{Li2016Investigation}.
\par
\begin{figure}
\centering
\includegraphics[width=8.5 cm]{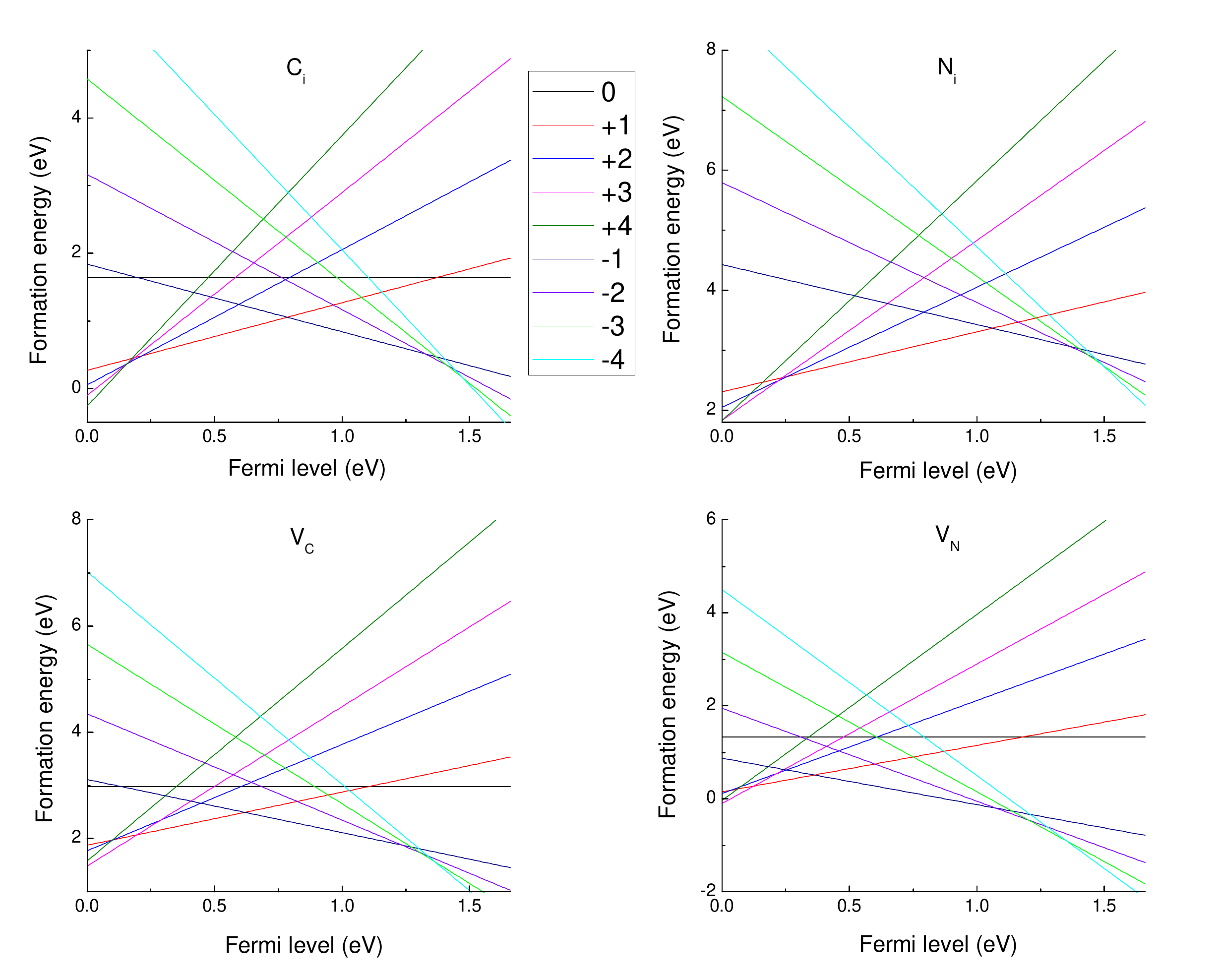}
\caption{The formation energies of native defects ($\rm C_i$, $\rm V_C$, $\rm N_i$, and $\rm V_N$) versus Fermi level under carbon rich limit.}
\label{figure7}
\end{figure}
\section{Conclusions}
In summary, we perform a detailed first principles analysis of NHG. It is found from density of states of NHG that the valence band is mainly composed of C 2s, 2p and N 2s, 2p. The C 2s, 2p hybridizes with N 2s, 2p at valence band. The charge density is distributed between C and N, so covalent bonds are formed between N and C atoms. Interlayer stacking behavior is studied by binding energy, and the most stable A-F stacking structure is obtained. It is found that the  binding energy is -26.38 meV/atom for A-F stacking, which is higher than -57.53 meV/atom of the graphene. This result indicates that the interlayer interaction is weak for NHG, so the monolayer NHG is easy to form. The band structures suggest that $\rm NHG$ is a direct band gap semiconductor, and its value decreases with the increase of the number of $\rm NHG$ layers. For monolayer NHG, the lattice strain can modulate the band gap, thus affecting the carrier transport properties. The direct band gap calculated by GGA-PBE functional changes in the range from 1.297 eV - 1.708 eV as the strain applied to the monolayer NHG. By analyzing the effect of strain on bond length, we find that NHG has good elasticity. The native defects of $\rm C_i$, $\rm V_C$, $\rm N_i$, and $\rm V_N$ are unstable in the neutral charge state. $\rm C_i$ and $\rm N_i$ easy to form in hole. These defects are easily ionized and act as carrier traps and fixed charge, which can scatter carrier and shift the threshold voltage.
\section{Acknowledgements}
The work is supported by the Science Foundation from Education Department of Liaoning Province, China (Grant Nos. LF2017001 and LQ2017005).

\end{document}